# Synchronous Spatial Oscillation of Electron- and Mn-Spin Polarizations in Dilute-Magnetic-Semiconductor Quantum Wells under Spin-Orbit Effective Magnetic Fields

Takuma TSUCHIYA

*Division of Applied Physics, Faculty of Engineering, Hokkaido University (Hokudai)*
*Kita 13 Nishi 8, Kita-ku, Sapporo 060-8628, Japan*

In semiconductors, spin-orbit effective magnetic fields, i.e., the Rashba and Dresselhaus fields, are used to control electron-spin polarization. This operation, however, destroys the electron-spin coherence, and the spin polarization is limited to the vicinity of a ferromagnetic source electrode. In this paper, we propose the use of dilute magnetic semiconductors to improve the coherence of spatially oscillating electron-spin polarization. In dilute magnetic semiconductors, the electron-spin polarization near the source electrode dynamically induces the local spin polarization of magnetic impurities through s-d spin-flip scattering. This impurity-spin polarization improves, in turn, the coherence of the electron-spin polarization, and this improved electron-spin polarization induces impurity-spin polarization farther in the adjacent region. Because of this positive feedback, the coherent and synchronized spatial oscillations of electron- and impurity-spin polarizations grow cooperatively. A numerical calculation for a CdMnTe quantum well demonstrates the validity of this mechanism.

1. Introduction

Controlling the spatial distribution of electron-spin polarization, i.e., the local average of the spin vectors of many electrons, without deteriorating electron-spin coherence is quite important for spintronics device applications.[1] However, it is not an easy task. In some devices, such as spin field-effect transistors (spin-FETs),[2] electron spins, which are injected into the two-dimensional (2D) channel from a ferromagnetic source electrode, are controlled through electron-spin precession caused by the Rashba[3,4] and Dresselhaus[5] effective magnetic fields, both originating from the atomic spin-orbit interaction. However, because these spin-orbit fields and the resulting electron-spin precession depend on the electron wave vector, the spin coherence length is limited through the D'yakonov-Perel' (DP) relaxation mechanism,[6] and the electrons-spin polarization is localized near the source electrode. Therefore, it is worthwhile to search for methods to overcome the DP spin relaxation.

Dilute magnetic semiconductors (DMSs), investigated actively for over thirty years,[7] have been considered inappropriate for coherent electron-spin transport. In these materials, various fascinating phenomena, such as carrier-induced ferromagnetism, giant Zeeman effect, and spin polaron effect,[8] are caused by the s-d exchange interaction between the spins of conduction s-electrons and localized d-electrons of magnetic impurities. Regrettably, this s-d interaction destroys the electron-spin coherence because of the s-d spin-flip scattering caused by the s-d interaction. Turning our attention to the polarization of impurity spins, however, we note a possibility to improve the electron-spin coherence. It was pointed out in ref. 9 that the impurity spins are polarized dynamically by s-d spin-flip scattering under electron-spin polarization, and this dynamical impurity-spin polarization could induce ferromagnetic ordering even under the small splitting of the up-spin and down-spin quasi-Fermi levels for carriers. This is a result of the giant Zeeman splitting for electrons under the induced impurity-spin polarization. If the electron-spin polarization oscillates spatially, the induced impurity-spin polarization also oscillates. In this case, the electron spins are not always parallel to the impurity-spin polarization, and the electron-spin precession is expected to be changed by the effective magnetic field originating from the impurity-spin polarization. It is quite desirable if this change in electron-spin precession improves the electron-spin coherence.

In this paper, we propose a possible mechanism to improve the spatial electron-spin coherence by DMSs and demonstrate its validity through a numerical calculation for the spin transport of conduction electrons injected from a ferromagnetic source electrode into a 5 nm $Cd_{0.99}Mn_{0.01}Te$ quantum well (QW). This paper is organized as follows. In §2, we propose a possible mechanism to overcome the DP spin relaxation for conduction electrons. In §3, a method of numerical calculation is explained. Numerical results are shown in §4, and §5 is devoted to a summary. In Appendix, we give a brief review for the electron-spin polarization in nonmagnetic quantum wells.

2. **Dynamical Mn-spin polarization and improvement of electron-spin coherence**

We consider the transport of spin-polarized conduction electrons in dilute-magnetic QWs. As is shown schematically in Fig. 1, electrons are injected from a ferromagnetic source electrode into the QW at $(x, y) = (0, \text{arbitrary})$. Their spins are assumed to be polarized along the $z$-direction. When the electrons are injected, the direction $\theta_k$ of the 2D wave vector $\mathbf{k}_\parallel = (k_x, k_y) = k_\parallel \cdot (\cos\theta_k, \sin\theta_k)$ of each electron is random, and the 2D wave number $k_\parallel$ is distributed in accordance with the Fermi distribution function. In the QW, the electric field $E_x$ between a source electrode and a drain electrode is applied along the $x$-axis. In the present paper, we ignore the Dresselhaus field for simplicity and take into account only the Rashba field for the spin-orbit effective magnetic field. Because the system has a translational symmetry along the $y$-axis, the electron-spin polarization is uniform along the $y$-axis and depends only on $x$. We do not consider the details of the ferromagnetic electrode, because the details of the spin injection are beyond the scope of this study.

For electron transport, we assume that the state of an electron along the QW plane is given by well-defined position and momentum. For this assumption, it is necessary that the uncertainties of the position and momentum are much smaller than the electron mean free path and momentum, respectively. This condition is known to be satisfied usually in semiconductors, and this assumption is widely employed in the Monte Carlo simulations of electron transport in semiconductors.[10,11] Thus, the state of an electron is specified by confinement wave function $\varphi(z)$, 2D position $\mathbf{r}_\parallel = (x, y)$, 2D wave vector $\mathbf{k}_\parallel$, and spin vector $\mathbf{s} = (s_x, s_y, s_z)$. Under the effective-mass approximation, the 2D electron velocity is given by $\mathbf{v}_\parallel = \hbar \mathbf{k}_\parallel / m^*$, where $m^*$ is the effective mass of the conduction band. The electrons suffer momentum scatterings by phonons and impurities, as is schematically shown in Fig. 1.

The precession of an electron spin is determined, in general, by the precession equation

$$\frac{d\mathbf{s}}{dt} = \mathbf{\Omega}_{pr} \times \mathbf{s}, \tag{1}$$

where $\mathbf{\Omega}_{pr}$ is the precession vector.[12] For the Rashba effective magnetic field induced by the applied electric field $E_z$ along the growth direction, the precession vector is given by

$$\mathbf{\Omega}_{\text{Rashba}}\left(\mathbf{k}_{\|}\right) = -\gamma \mathbf{B}_{\text{Rashba}}\left(\mathbf{k}_{\|}\right) = \frac{2\alpha_{\text{Rashba}}}{\hbar}\left(k_y, -k_x, 0\right), \qquad (2)$$

where $\mathbf{B}_{\text{Rashba}}$ is the Rashba field, $\gamma$ the gyromagnetic ratio of a conduction electron, and $\alpha_{\text{Rashba}}$ ($\propto E_z$) the Rashba coefficient.[3,4] A schematic illustration of electron-spin precession is shown in Fig. 2(a). Because the direction of $\mathbf{\Omega}_{\text{Rashba}}$ is along the QW and perpendicular to $\mathbf{k}_{\|}$, the electron spin rotates in the plane including $\mathbf{k}_{\|}$ and the $z$-axis. This $\mathbf{k}_{\|}$ dependence of the precession and the momentum scatterings due to acoustic and optical phonons, and nonmagnetic impurities cause the DP spin relaxation. As a result, the spin polarization $\langle \mathbf{s}(x) \rangle$, or the local average of many electron spins in a small spatial region around $x$, is localized near the source electrode as will be shown in Fig. 5(c) in Appendix.

Let us consider the effects of the s-d exchange interaction on electron-spin polarization. This interaction causes two phenomena important for the present research: the dynamical impurity-spin polarization originating from the s-d spin-flip scattering[9] and the electron-spin precession due to this impurity-spin polarization. The s-d Hamiltonian[7] is given by

$$H_{\text{s-d}} = -a_{\text{s-d}} \sum_i \hat{\mathbf{S}}_i^{\text{Mn}} \cdot \hat{\mathbf{s}} \delta\left(\mathbf{r} - \mathbf{R}_i^{\text{Mn}}\right), \qquad (3)$$

where $\hat{\mathbf{s}}$ and $\hat{\mathbf{S}}_i^{\text{Mn}}$ are the spin operators of a conduction s-electron and the $i$-th Mn impurity, and $\mathbf{r} = (x, y, z)$ and $\mathbf{R}_i^{\text{Mn}} = (X_i, Y_i, Z_i)$ are their coordinates, respectively. $a_{\text{s-d}}$ is the s-d coupling constant. Considering the $z_p$-axis as a principal axis for spins and using the rising and lowering operators $\hat{s}_\pm = \hat{s}_{x_p} \pm i\hat{s}_{y_p}$ and $\hat{S}_{i\pm}^{\text{Mn}} = \hat{S}_{i,x_p}^{\text{Mn}} \pm i\hat{S}_{i,y_p}^{\text{Mn}}$, we obtain

$$H_{\text{s-d}} = -a_{\text{s-d}} \sum_i \left[ \frac{1}{2}\left( \hat{S}^{\text{Mn}}_{i+}\hat{s}_- + \hat{S}^{\text{Mn}}_{i-}\hat{s}_+ \right) + \hat{S}^{\text{Mn}}_{iz_{\text{p}}}\hat{s}_{z_{\text{p}}} \right] \delta\left(\mathbf{r} - \mathbf{R}^{\text{Mn}}_i\right). \quad (4)$$

In this equation, the term $\hat{S}^{\text{Mn}}_{i+}\hat{s}_- + \hat{S}^{\text{Mn}}_{i-}\hat{s}_+$ gives the s-d spin-flip scattering. Because this scattering transfers electron spins into Mn spins, the Mn-spin polarization $\langle \mathbf{S}^{\text{Mn}}(x,t) \rangle$, or the local average of $\mathbf{S}^{\text{Mn}}$ in a small spatial region around $x$, is induced dynamically under the electron-spin polarization. This dynamical Mn-spin polarization[9] is similar to the Overhauser effect between spin-polarized electrons and nuclei.[12,13]

The other term $\hat{S}^{\text{Mn}}_{iz_{\text{p}}}\hat{s}_{z_{\text{p}}}$ in eq. (4) causes the energy splitting between the spin-up and spin-down states of electrons under a finite $\langle \mathbf{S}^{\text{Mn}}(x,t) \rangle$, resulting in electron-spin precession for the mixed spin state. Because each electron is assumed to form a wave packet, it is reasonable to consider the electron-spin precession vector due to the Mn-spin polarization to be determined by $\langle \mathbf{S}^{\text{Mn}}(x,t) \rangle$. The precession vector is given by

$$\mathbf{\Omega}_{\text{Mn}}(x,t) = -\gamma \mathbf{B}_{\text{Mn}}(x,t) \simeq -n_{\text{Mn}} \frac{a_{\text{s-d}}}{\hbar} \langle \mathbf{S}^{\text{Mn}}(x,t) \rangle, \quad (5)$$

where $n_{\text{Mn}}$ is the Mn density. The total electron-precession vector is given simply by the sum of eqs. (2) and (5) as

$$\mathbf{\Omega}_{\text{pr}}(\mathbf{k}_\parallel, x, t) = \mathbf{\Omega}_{\text{Rashba}}(\mathbf{k}_\parallel) + \mathbf{\Omega}_{\text{Mn}}(x,t) = -\gamma \mathbf{B}_{\text{Rashba}}(\mathbf{k}_\parallel) - \gamma \mathbf{B}_{\text{Mn}}(x,t). \quad (6)$$

If the condition $|\mathbf{B}_{\text{Mn}}| \gg |\mathbf{B}_{\text{Rashba}}|$ is satisfied, $\mathbf{B}_{\text{Mn}}$ dominates the electron-spin precession, and the spatial electron-spin coherence is expected to be improved. Let us consider the behavior of an electron spin vector under $\mathbf{B}_{\text{Mn}}(x)$. For example, we consider $\mathbf{B}_{\text{Mn}}(x)$ to be proportional to $\langle \mathbf{S}^{\text{Mn}}(x) \rangle$ shown in Fig. 3(b2), which will be obtained numerically in §4. This $\langle \mathbf{S}^{\text{Mn}}(x) \rangle$ is induced by electron-spin polarization similarly to $\langle \mathbf{s}(x) \rangle$ shown in Fig. 3(a1) and is almost parallel to it. At the source edge, the injected electron spins are parallel to $\mathbf{B}_{\mathrm{Mn}}(x=0)$, and the direction of $\mathbf{B}_{\mathrm{Mn}}(x)$ changes gradually as the electrons move in the channel. When $\mathbf{B}_{\mathrm{Mn}}$ is strong enough, the electron spin almost follows $\mathbf{B}_{\mathrm{Mn}}(x)$ adiabatically regardless of $\mathbf{B}_{\mathrm{Rashba}}(\mathbf{k}_\parallel)$, as is schematically shown in Fig. 2(b). This mechanism is robust against the momentum scatterings of electrons, because the change in $\mathbf{B}_{\mathrm{Rashba}}(\mathbf{k}_\parallel)$ is negligible for $|\mathbf{B}_{\mathrm{Mn}}| \gg |\mathbf{B}_{\mathrm{Rashba}}|$. This is in contrast to electron-spin precession only by $\mathbf{B}_{\mathrm{Rashba}}$, as schematically shown in Fig. 2(a), where the precession depends on $\theta_k$. Therefore, all electron spins at $x$ are almost parallel to $\mathbf{B}_{\mathrm{Mn}}(x)$, and the spatial electron-spin coherence is expected to be improved. This improved spatial electron-spin coherence elongates the region of the induced Mn-spin polarization in turn. This extended Mn-spin polarization further improves the spatial electron-spin coherence. This positive feedback between the electron- and Mn-spin polarizations elongates their coherence lengths successively.

The condition $|\mathbf{B}_{\mathrm{Mn}}| \gg |\mathbf{B}_{\mathrm{Rashba}}|$ necessary for the present mechanism is easily satisfied in typical II-VI DMS QWs on a time scale of 10 ns. To perform an order estimation, we consider a 10 nm $Cd_{0.99}Mn_{0.01}Te$ QW with the electron sheet density $N_S = 10^{12}$ cm$^{-2}$. The electron-spin splitting $\Delta_{\mathrm{Mn}} = \hbar|\mathbf{\Omega}_{\mathrm{Mn}}| = N_0 a_{\mathrm{s\text{-}d}} x_{\mathrm{Mn}} |\langle \mathbf{S}^{\mathrm{Mn}} \rangle|$ under fully polarized Mn spins, or $|\langle \mathbf{S}^{\mathrm{Mn}} \rangle| = 5/2$, is estimated to be 5.5 meV, where $N_0 = 4/a^3$ is the density of cation sites with $a$ being the lattice constant, $x_{\mathrm{Mn}}(=0.01)$ the Mn mole fraction, and $N_0 a_{\mathrm{s\text{-}d}} = 220$ meV for $Cd_{1-x_{\mathrm{Mn}}}Mn_{x_{\mathrm{Mn}}}Te$.[7] This exceeds the spin splitting by the Rashba field $\Delta_{\text{Rashba}} = 2\alpha_{\text{Rashba}} k_F = 0.6$ meV for an electron at the 2D Fermi surface for $N_S = 10^{12}$ cm$^{-2}$ or $k_F \simeq 0.03$ Å$^{-1}$, and $\alpha_{\text{Rashba}} = 10$ meVÅ for example. This Rashba field corresponds to the applied gate electric field $E_z \simeq 18$ mV/nm estimated from

$$\alpha_{\text{Rashba}} = \frac{\hbar^2}{2m^*} \times \frac{\Delta_{\text{SO}}(2E_g + \Delta_{\text{SO}})}{E_g(E_g + \Delta_{\text{SO}})(3E_g + 2\Delta_{\text{SO}})} eE_z, \tag{7}$$

where $E_g = 1.606$ eV is the band-gap energy, $\Delta_{\text{SO}} = 0.8$ eV the spin-orbit splitting in the valence band,[14] and $e$ the elementary charge.

The time scale of the growth of the Mn-spin polarization is obtained to be 10 ns from $\tau_e^{\text{sf}}$, or the s-d spin-flip scattering time of electrons, which is estimated to be $\tau_e^{\text{sf}} \sim 100$ ps for the present QW.[15] Since the density of Mn, $n_{\text{Mn}} = N_0 x_{\text{Mn}} \simeq 147 \times 10^{18}$ cm$^{-3}$, is about 100 times larger than the electron volume density $n_e = N_S/d = 10^{18}$ cm$^{-3}$ for the well thickness $d = 10$ nm, the order of Mn spin-flip scattering time is $\tau_{\text{Mn}} = n_{\text{Mn}} \tau_e / n_e \sim 10$ ns. Mn-spin scattering mechanisms other than the s-d scattering, which originate, for example, from phonon scatterings,[8] can be ignored for low temperatures, because their time scale, 0.1 ms at 4.2 K,[16-23] is much longer than the s-d scattering time.

In addition to the s-d spin-flip scattering, the spatial Mn-spin fluctuation causes the electron-spin relaxation. Although the experimental relaxation time[24] of $\tau_e^{\text{fluct}} \simeq 10$ ps was observed for 8 nm Cd$_{0.955}$Mn$_{0.045}$Te QWs with $N_S = 3 \times 10^{10}$ and $7 \times 10^{10}$ cm$^{-2}$, we expect that this type of relaxation can be ignored under the in-plane electric field $E_x$, particularly because of the following three reasons: First, $\tau_e^{\text{fluct}}$ in the present Cd$_{0.99}$Mn$_{0.01}$Te QW is expected to be about 45 ps, because $\tau_e^{\text{fluct}} \propto 1/x_{\text{Mn}}$.[8] Second, it takes only 5 ps, or one order shorter than the relaxation time, for electrons to drift 1 μm under $|E_x| = 1$ kV/cm for example, when the electron mobility $\mu = 2 \times 10^4$ cm$^2$/Vs, which is obtained by the Monte Carlo calculation in §4. Actually, the higher mobility $\mu = 2.6 \times 10^5$ cm$^2$/Vs was observed experimentally for the 20 nm CdTe QW at 0.6 K.[25] Third, the above spin relaxations are expected to be relieved with increasing Mn-spin polarization, because the Mn-spin fluctuation is suppressed.

## 3. Numerical Method

In order to demonstrate the validity of the above-proposed mechanism, we perform a numerical calculation. In this calculation, processes essential for the present mechanism, i.e., electron transport with momentum scatterings, dynamical Mn-spin polarization, and electron-spin precession due to $\mathbf{B}_{Mn}$ and $\mathbf{B}_{Rashba}$, are considered. For the wave function of electrons confined in the QW, we take into account only the ground subband for simplicity and employ the infinite-barrier approximation. Then, the electron wave function along the $z$-direction is given by $\varphi(z) = \sqrt{2/d} \sin(\pi z/d)$, in which we have ignored the modification by $E_z$ for simplicity.

To simulate electron transport, we employ the Monte Carlo method.[10,11,26-28] *Virtual* electrons, spin-polarized along the $z$-axis, are injected into the channel from the ferromagnetic source electrode. The energy $\varepsilon$ of each electron is given by the Monte Carlo method in accordance with the Fermi distribution. The 2D momentum of this electron is given by $\mathbf{k}_\parallel = \sqrt{2m^*\varepsilon}/\hbar \times (\cos\theta_k, \sin\theta_k)$, where $\theta_k$ is given by a uniform pseudo-random number between $-\pi/2$ and $\pi/2$. These electrons are accelerated by an electric field $E_x$ along the channel and scattered sometimes by acoustic and optical phonons, nonmagnetic impurities, and magnetic Mn impurities. The probabilities of these scatterings are estimated through Fermi's golden rule in the usual manner, and the timing of each scattering event is determined also by the Monte Carlo method.[11] Because the details of nonmagnetic-impurity scattering are not important for the present calculation, we assume the impurity scattering time to be 15 ps. Since the anomalous and spin Hall effects are beyond the scope of this study, processes related to them, such as skew scattering and side jump,[29] are not taken into account. The source and drain electrodes are assumed to absorb all incoming virtual electrons. When a virtual electron is absorbed by these electrodes, a new virtual electron is emitted from the source electrode. The drain electrode is at $x = 6$ μm for the present calculation.

We estimate the rate of the s-d spin-flip scattering between electrons and Mn spins under the assumption that the Mn spins in $Cd_{0.99}Mn_{0.01}Te$ are isolated. For this Mn concentration, the strong anti-ferromagnetic interaction between nearest-neighbor Mn impurities can be ignored,[7] because the probability for a Mn impurity to have at least one nearest-neighbor Mn impurity is $1-(0.99)^4 \simeq 0.04$. The probability of spin-flip scattering from spin-up to spin-down electron states is given by Fermi's golden rule for the s-d Hamiltonian, eq. (4), as[15]

$$W_{sf} = \sum_i a_{s-d}^2 \frac{\pi^2 m}{\hbar^3} \left(\frac{L}{2\pi}\right)^2 \frac{|\varphi^*(Z_i)\varphi(Z_i)|^2}{L^4} F(m_i) \Theta\left(\frac{\hbar^2 k_\parallel^2}{2m^*} + E_+^{s-d} - E_-^{s-d}\right),$$

$$F(m_i) = j_i(j_i+1) - m_i(m_i+1) = \frac{5}{2}\left(\frac{5}{2}+1\right) - m_i(m_i+1),$$

(8)

where $L^2$ is the area of the QW, $\Theta$ Heaviside's step function, $E_+^{s-d}$ and $E_-^{s-d}$ are the spin energies originating from the s-d interaction for the spin-up and spin-down states, $j_i(=5/2)$ is the length of the Mn spin, and $m_i$ is the Mn-spin component along the direction of the electron spin. We ignore the Pauli exclusion principle for the final-state occupation of electrons for simplicity. This is reasonable for sufficiently spin-polarized electrons, which are important for the present mechanism, because the final spin-down states are almost unoccupied. Although the rate of this spin-flip scattering due to individual Mn impurities depends in reality on $Z_i$ and $F(m_i)$, we consider, for simplicity, an average over Mn impurities in the present calculation. Assuming that the Mn concentration $n_{\text{Mn}}$ is uniform in the QW, we obtain[15]

$$\sum_i |\varphi^*(Z_i)\varphi(Z_i)|^2 \simeq n_{\text{Mn}} dL^2 \int_0^d \varphi^4(z) \cdot \frac{1}{d} dz$$
$$= \frac{4n_{\text{Mn}} dL^2}{d^3} \int_0^d \sin^4\left(\frac{\pi}{d} z\right) = \frac{4n_{\text{Mn}} L^2}{d^2} \frac{6d}{16} = \frac{3n_{\text{Mn}} L^2}{2d}. \tag{9}$$

For the factor $F$ in eq. (8), we assume the phenomenological form

$$F_{\text{av}}(\langle m(x_j)\rangle) = \begin{cases} \dfrac{35}{6}\cos\left(\dfrac{\langle m(x_j)\rangle \pi}{5}\right) & \text{for } \langle m(x_j)\rangle \geq 0 \\ \dfrac{35}{6} & \text{for } \langle m(x_j)\rangle < 0 \end{cases}, \tag{10}$$

where $\langle m(x_j)\rangle$ is an average of $m_i$ in a small grid between $x = j\Delta x$ and $(j+1)\Delta x$, where $j = 0, 1, 2, \cdots$ and $\Delta x$ is the width of grids. This form reproduces $F_{\text{av}}(\langle m(x_j)\rangle = 0) = 35/6$ for a random distribution of $m_i$, and $F_{\text{av}}(5/2) = 0$ and $F_{\text{av}}(-5/2) = 5$ for fully polarized Mn spins.[15] Hence, we obtain

$$W_{\text{sf}}(x_j) \simeq \frac{3a_{\text{s-d}}^2 m n_{\text{Mn}}}{8\hbar^3 d} F_{\text{av}}(\langle m(x_j)\rangle) \Theta\left(\frac{\hbar^2 k_\parallel^2}{2m^*} + E_+^{\text{s-d}} - E_-^{\text{s-d}}\right). \tag{11}$$

In order to estimate the time evolution of $\langle \mathbf{S}_{\text{Mn}}(x,t)\rangle$, we use the total change of electron spin $\Delta\langle \mathbf{s}(x_j,t_k)\rangle$, obtained by the Monte Carlo process, in the space grid and a time grid between $t = k\Delta t$ and $(k+1)\Delta t$ with the integer $k$. Because the total angular momentum is conserved in the s-d spin flip, we obtain

$$n_e \Delta \langle \mathbf{s}(x_j, t_k) \rangle + n_{Mn} \Delta \langle \mathbf{S}_{Mn}(x_j, t_k) \rangle = 0, \tag{12}$$

where $\Delta \langle \mathbf{S}_{Mn}(x_j, t_k) \rangle$ is the change in $\langle \mathbf{S}_{Mn}(x_j, t) \rangle$. Taking into account the spin-lattice relaxation time $\tau_{Mn}^{lattice}$ caused by phonon scatterings,[8] we obtain the time evolution of the Mn-spin polarization by

$$\langle \mathbf{S}_{Mn}(x_j, t_{k+1}) \rangle = \langle \mathbf{S}_{Mn}(x_j, t_k) \rangle - \Delta \langle \mathbf{s}(x_j, t_k) \rangle \frac{n_e}{n_{Mn}} - \frac{\left[ \langle \mathbf{S}_{Mn}(x_j, t_k) \rangle - \mathbf{S}_{Mn}^0 (\langle \mathbf{s}(x_j, t_k) \rangle, T) \right] \Delta t}{\tau_{Mn}^{lattice}}, \tag{13}$$

where $\mathbf{S}_{Mn}^0(\langle \mathbf{s}(x_j, t_k) \rangle, T)$ is the Brillouin function of Mn spins for the effective magnetic field originating from $\langle \mathbf{s}(x_j, t_k) \rangle$ at the temperature $T$. Actually, $\mathbf{S}_{Mn}^0$ is negligible, because the Mn-spin splitting is one order of magnitude smaller than the thermal energy even at 4.2 K. We employ $\tau_{Mn}^{lattice} = 0.1$ ms at $T = 4.2$ K.[16-23] Although the actual s-d spin-flip scattering rate of Mn depends on $Z_i$, we have ignored this dependence, in accordance with the averaging over Mn impurities employed above. The spin diffusion of Mn is also ignored, because the experimental diffusion constant is small for similar II-VI DMSs; $K_{diff} = 7 \times 10^{-8}$ cm$^2$/s at 1.8 K for Zn$_{0.99}$Mn$_{0.01}$Se.[30] The precession of Mn spins due to the effective magnetic field caused by $\langle \mathbf{s}(x,t) \rangle$ is neglected, because $\langle \mathbf{S}_{Mn}(x,t) \rangle$ is almost parallel to $\langle \mathbf{s}(x,t) \rangle$ and the precession frequency is low.

It should be noted that $\langle \mathbf{S}_{Mn}(x,t) \rangle$ is overestimated in some degree because of the averaging. However, this overestimation is expected not to change the numerical results much. In reality, the average of the Mn spins should be given in equilibrium by

$$\langle S_{Mn}^z \rangle = \frac{\sum_{S_{Mn}^z=-5/2}^{5/2} S_{Mn}^z \cdot \left(p_e^{+1/2}/p_e^{-1/2}\right)^{S_{Mn}^z}}{\sum_{S_{Mn}^z{}'=-5/2}^{5/2} \left(p_e^{+1/2}/p_e^{-1/2}\right)^{S_{Mn}^z{}'}}, \qquad (14)$$

because the Mn-spin population $p_{Mn}^{S_{Mn}^z}$ satisfies $p_e^{+1/2} p_{Mn}^{S_{Mn}^z} = p_e^{-1/2} p_{Mn}^{S_{Mn}^z+1}$, where $p_e^{+1/2}$ and $p_e^{-1/2} = 1 - p_e^{+1/2}$ are the electron-spin populations for $s_z = +1/2$ and $-1/2$, respectively. This equation gives $\langle S_{Mn}^z \rangle \simeq 1.5$ and 2, which correspond to $\Delta_{Mn} \simeq 3.3$ and 4.4 meV, for $\langle s^z \rangle = \left(p_e^{+1/2} - p_e^{-1/2}\right)/2 \simeq 0.15$ and 0.25, respectively, for example. Thus, the electron-spin splitting $\Delta_{Mn}$ due to Mn-spin polarization is much larger than that by the Rashba field, $\Delta_{Rashba} = 0.6$ meV, even under a weak electron-spin polarization. Therefore, the overestimation changes the numerical results much only in regions where the electron-spin polarization is quite weak. The above estimation also indicates that the present mechanism is valid even for the insufficient spin polarization of injected electrons.

The spin precession of individual electrons is calculated numerically from eqs. (1), (2), (5), and (6). $\mathbf{\Omega}_{Mn}$ is assumed to be constant within a time and space grid. The widths of the time and space grids used in the present calculation are $\Delta x = 0.03$ μm and $\Delta t = 1$ ps, respectively. We have assumed $\langle \mathbf{S}_{Mn}(x,t<0) \rangle = 0$, and the s-d interaction is switched on at $t=0$. The number of virtual electrons used in the calculation is 50,000. The parameters used in the present calculation are shown in Table I.

## 4. Numerical Results

In Fig. 3, we show the electron- and Mn-spin polarizations, $\langle \mathbf{s}(x) \rangle$ and $\langle \mathbf{S}_{\mathrm{Mn}}(x) \rangle$, in a 5 nm $Cd_{0.99}Mn_{0.01}Te$ QW with the electron sheet density $N_\mathrm{S} = 10^{12}$ cm$^{-2}$, the Rashba coefficient $\alpha_{\mathrm{Rashba}} = 10$ meVÅ, the in-plane electric field $E_x = -1$ kV/cm, and the temperature $T = 4.2$ K as a function of the distance $x$ from the source electrode for the elapsed times $t = 0$, 5, and 40 ns. At $t = 0$, Mn spins are not polarized or $\langle \mathbf{S}_{\mathrm{Mn}}(x) \rangle = 0$, as shown in Fig. 3(a2), and the electron-spin precession is caused only by the Rashba field. In Figs. 4(a1)-4(a5), we show the components of individual electron spins as a function of $x$ for the constant $\mathbf{k}_\parallel = k_\parallel \cdot (\cos\theta_k, \sin\theta_k)$ with $\theta_k = 0$, $\pm\pi/6$, and $\pm\pi/3$, and $k_\parallel = k_\mathrm{F}/2 = 1.77 \times 10^6$ cm$^{-1}$, where $k_\mathrm{F}$ is the Fermi wave number for spin-polarized electrons with the sheet density $N_\mathrm{S} = 10^{12}$ cm$^{-2}$. Although it is found that these profiles depend on the direction $\theta_k$, they are independent of the wave number $k_\parallel$. This is because both $|\mathbf{v}_\parallel|$ and $|\mathbf{\Omega}_{\mathrm{Rashba}}(\mathbf{k}_\parallel)|$ are proportional to $k_\parallel$. Because of these $\theta_k$ dependence and momentum scatterings, $\langle \mathbf{s}(x) \rangle$ is damped within a half oscillation period and vanishes for $x \geq 2$ μm. This profile is essentially the same as the result for the nonmagnetic CdTe QW shown in Fig. 5(c) in Appendix, because the effect of the s-d spin-flip scattering on electron-spin polarization is weak.

For $t = 5$ ns, $\langle \mathbf{S}_{\mathrm{Mn}}(x) \rangle$ is partially polarized for $x < 1.5$ μm, as is shown in Fig. 3(b2). This profile is almost proportional to $\langle \mathbf{s}(x) \rangle$ at $t = 0$ in Fig. 3(a1). At the same time, the spatial coherence of $\langle \mathbf{s}(x) \rangle$ shown in Fig. 3(b1) is improved in this region. The profiles of individual electron spins under this $\langle \mathbf{S}_{Mn}(x) \rangle$ are shown in Figs. 4(b1) -4(b5). As has been expected, the electron spins are almost parallel to $\langle \mathbf{S}_{Mn}(x) \rangle$ regardless of $\mathbf{k}_\parallel$ for $x < 1.5$ μm, where Mn spins are polarized. On the contrary, the spin profile depends on $\mathbf{k}_\parallel$ for $x > 1.5$ μm. For $x < 1.5$ μm, a small oscillation of $s_y$ is found. This oscillation is due to a small tilt of $\mathbf{\Omega}_{pr} = \mathbf{\Omega}_{Mn} + \mathbf{\Omega}_{Rashba}$ with respect to the axis of $\mathbf{s}$. The period of this oscillation is much shorter than that by the Rashba field in Figs. 4(a1)-4(a5), because $|\mathbf{\Omega}_{pr}|$ at $t = 5$ ns for $x < 1.5$ μm is much larger than $|\mathbf{\Omega}_{Rashba}|$.

As is shown in Fig. 3(c2), $\langle \mathbf{S}_{Mn}(x) \rangle$ is polarized almost fully at $t = 40$ ns. At the same time, $\langle \mathbf{s}(x) \rangle$ in Fig. 3(c1) shows a clear oscillation synchronous with $\langle \mathbf{S}_{Mn}(x) \rangle$ in the entire region in the figure. This is explained by the behavior of individual electron spins shown in Figs. 4(c1)-4(c5), which is almost independent of $\mathbf{k}_\parallel$ and almost parallel to $\langle \mathbf{S}_{Mn}(x) \rangle$. These results clearly demonstrate that the present mechanism is valid for overcoming the DP spin relaxation. It should be noted that that both $\langle \mathbf{s}(x) \rangle$ and $\langle \mathbf{S}_{Mn}(x) \rangle$ have small $y$-components for larger $x$ values and the individual electron spins show an additional small and rapid oscillation pronounced for larger $\theta_k$ values. The origin of the latter oscillation is the same as that of the oscillation at $t = 5$ ns for $x < 1.5$ μm. The small $\langle S_{Mn,y} \rangle$-component is induced by $\langle s_y(x) \rangle$ in the past. Actually, we find in Figs. 4(b1)-4(b5) that $s_y(x)$ tends to have a positive $y$-component for $x > 2$ μm at $t = 5$ ns. Because $|\mathbf{\Omega}_{Mn}|$ is comparable to $|\mathbf{\Omega}_{Rashba}|$ around $x \approx 2$ μm, the precession vector $\boldsymbol{\Omega}_{\text{pr}} = \boldsymbol{\Omega}_{\text{Rashba}} + \boldsymbol{\Omega}_{\text{Mn}}$ tilts toward the $y$-direction. As a result, electron spins have an $s_y$-component.

## 5. Summary

In this paper, we have proposed a possible mechanism to overcome the D'yakonov-Perel' spin relaxation for conduction electrons and to improve the spatial coherence of spatially oscillating spin polarization as a result. In this mechanism, the polarization of magnetic impurities in dilute-magnetic semiconductors, induced dynamically by the s-d interaction between conduction electrons and the impurities, plays an important role. The effective magnetic field due to the impurity-spin polarization can be stronger than the Rashba and Dresselhaus fields, and dominate the electron-spin precession. Under this condition, all electron spins follow the impurity field, regardless of their wave vectors and trajectories, and the spatial electron-spin coherence is improved. Numerical calculations, in which the Rashba effective magnetic field, the s-d interaction, and electron transport have been considered, have demonstrated that the synchronized and spatially coherent oscillations of electron- and magnetic-impurity-spin polarizations grow cooperatively owing to the positive feedback between them.

**Appendix: Electron-Spin Transport in Non-Magnetic Quantum Wells**

In this Appendix, we discuss the effects of the in-plane electric field $E_x$ and momentum scatterings on the electron-spin polarization $\langle \mathbf{s}(x) \rangle$ under the spin-orbit effective magnetic fields in nonmagnetic QWs. This is a starting point of the present study. Although we ignore the Dresselhaus effective field for simplicity, the results are qualitatively the same even under the Dresselhaus field. We consider the system schematically shown in Fig. 1. The electron-spin polarization, which is along the $z$-direction at the source edge, rotates spatially because of the spin precession of individual electrons due to the Rashba field, and it is relaxed through the DP spin-relaxation mechanism.[6] The method of numerical calculation is the same as that explained in § 3, except that the s-d interaction is not included.

For the present discussion, the $\theta_k$ dependence of $\mathbf{s}(x)$, or the $x$ dependence of spin precession for each electron, discussed already in §4 and shown in Figs. 4(a1)-4(a5) is important. Although these figures are for the 5 nm CdMnTe QW, the results are essentially the same as those for CdTe QWs, because $\langle \mathbf{S}_{\text{Mn}}(x) \rangle = 0$ in both cases. In Fig. 5(a), a numerical result of $\langle \mathbf{s}(x) \rangle$ for the 5 nm CdTe QW is shown. In this calculation, we assume that $\mathbf{k}_\parallel$ for each electron is given randomly and is time-independent. This means that we assume $E_x = 0$ and ignore momentum scatterings for electrons. The electron-spin polarization rotates in the $s_x$-$s_z$ plane, in spite that $s_y$ is finite for individual electrons with $\theta_k \neq 0$. This is due to the cancellation of $s_y$ between electrons with opposite $\theta_k$ values. The amplitude $|\langle \mathbf{s}(x) \rangle|$ decreases rapidly for $x < 1$ μm and gradually for $x > 1$ μm because of the $\theta_k$ dependence of electron-spin precession.

In the case of a finite $E_x$, electrons are accelerated along the $x$-axis, and their 2D wave vector direction converges to $\theta_k \to 0$. As a result, spin relaxation is expected to be reduced. The result for $E_x = -1$ kV/cm is shown in Fig. 5(b). As is expected, the spatial electron-spin coherence is improved considerably. In reality, however, it is necessary to take into account momentum scatterings, which change the direction of $\mathbf{k}_\parallel$ of individual electrons frequently. As a result, the spin precession of each electron is randomized, and the spin-coherence length is strongly reduced, as is shown in Fig. 5(c). This result is essentially the same as that for the magnetic CdMnTe QW with $\langle \mathbf{S}_{\text{Mn}}(x) \rangle = 0$ shown in Fig. 3(a). Thus, the momentum scatterings for electrons accelerate the DP spin relaxation strongly.

Under a finite $E_x$, $\langle k_x \rangle$, or $k_x$ averaged over all conduction electrons, and the resulting spin splitting along the $y$-axis due to the Rashba field are expected to be finite. Thus, we might anticipate a finite $\langle s_y \rangle$. However, this is not the case. To discuss spin polarization, it is necessary to consider the entire electron distribution in the $\mathbf{k}_\parallel$-space. As is shown in Fig. 6(a), the 2D Fermi surfaces for electrons with spins along the $\pm y$-direction for $E_x = 0$ at absolute zero are circular. This is because the dispersion relations for both spin states are parabolic and isotropic with respect to their bottoms at $O_u$ and $O_d$, even under the Rashba field. Since the areas of these circles are the same, the numbers of spin-up and down electrons are equal, and no spin polarization arises. Under a finite $E_x$, these circles are shifted along the $k_x$-axis, but their shapes are not changed within the Ohmic conduction regime, as is schematically shown in Fig. 6(b).[31] Although the circles may be warped in the non-Ohmic regime, the numbers of spin-up and spin-down electrons are expected to be the same. As a result, spin polarization is not induced by the Rashba field even under a finite $E_x$. This explanation is also valid for the Dresselhaus field, though the Fermi surfaces are not circular.

30) A. Maksimov, D. R. Yakovlev, J. Debus, I. I. Tartakovskii, A. Waag, G. Karczewski, T. Wojtowicz, J. Kossut, and M. Bayer: Phys. Rev. B **82** (2010) 035211.

31) C. Kittel: *Introduction to Solid State Physics* (Wiley, Hoboken, NJ, 2005) 8th ed.

32) D. L. Rode: Phys. Rev. B **2** (1970) 4036.
22

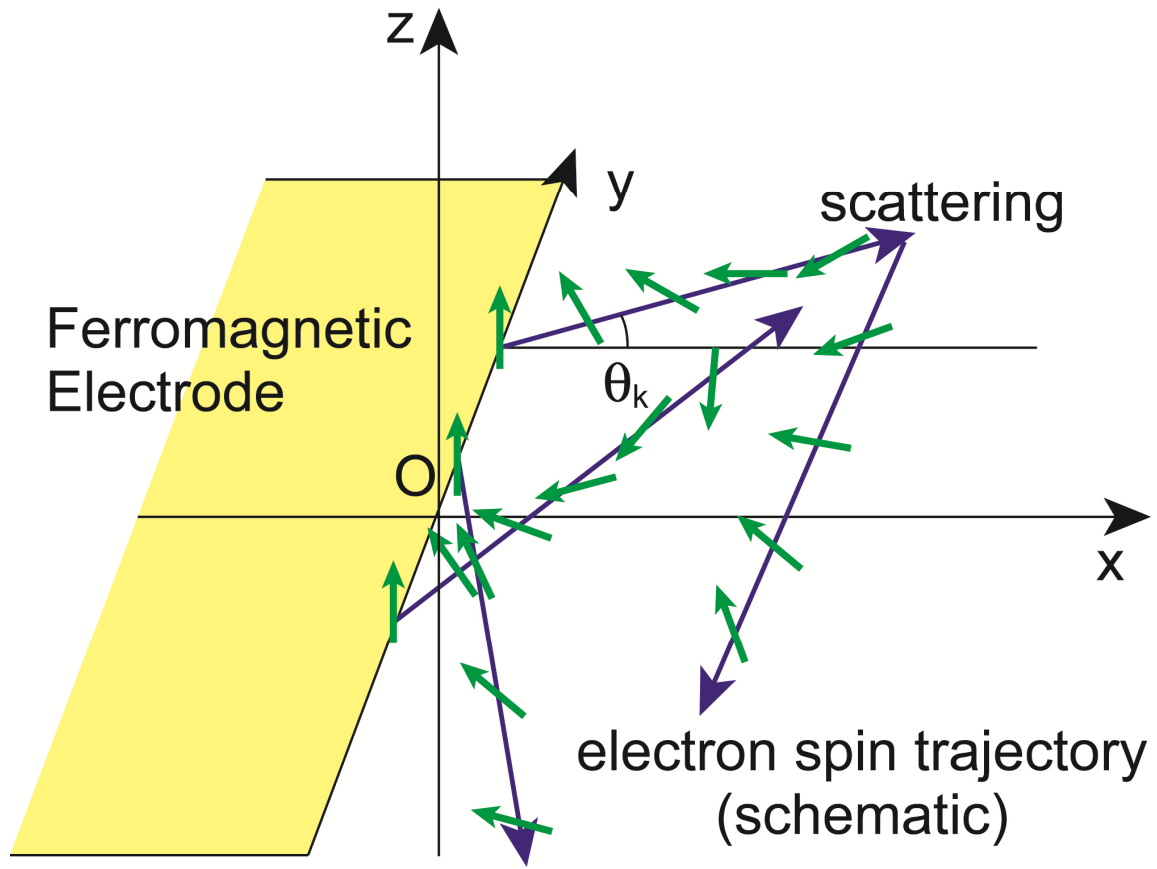

Fig. 1. (Color online) Schematic illustration of electron transport with momentum scatterings and electron-spin precession under spin-orbit effective magnetic fields.

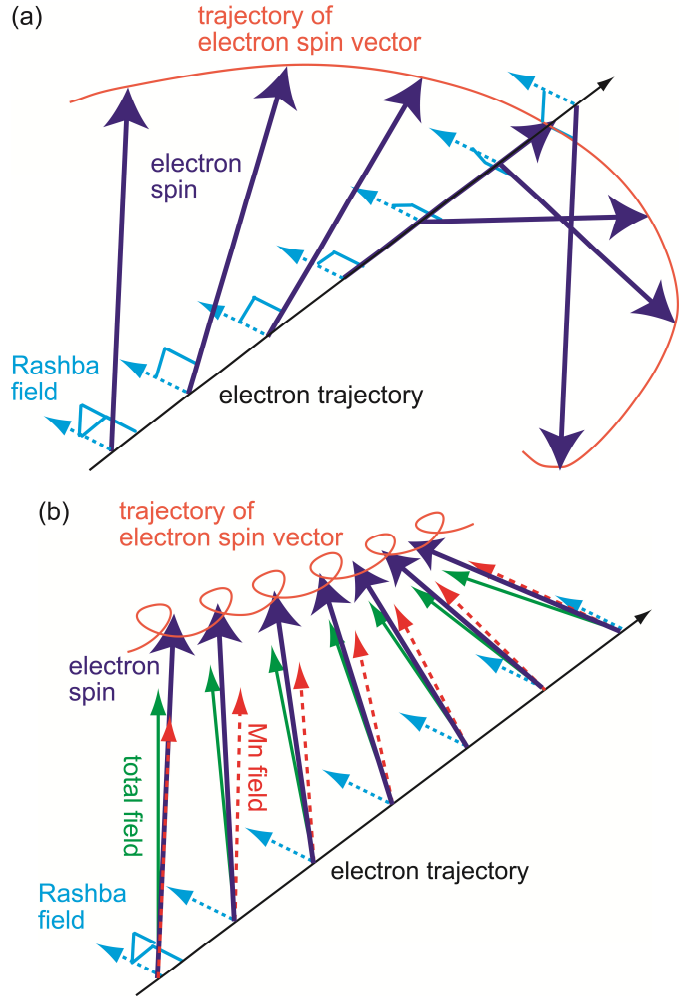

Fig. 2. (Color online) Schematic illustration of spin precession of an electron (a) without and (b) with strong Mn-spin polarization. (a) Without Mn-spin polarization, the axis of the electron-spin precession is parallel to the Rashba field. (b) Under the Mn-spin polarization, the electron spin follows the Mn-field adiabatically. Because of the small difference between the direction of the electron spin and the total effective magnetic field, a fine oscillation around the total field emerges. For visibility, the directions of the Rashba and Mn fields are made opposite to the system of the present numerical calculation.

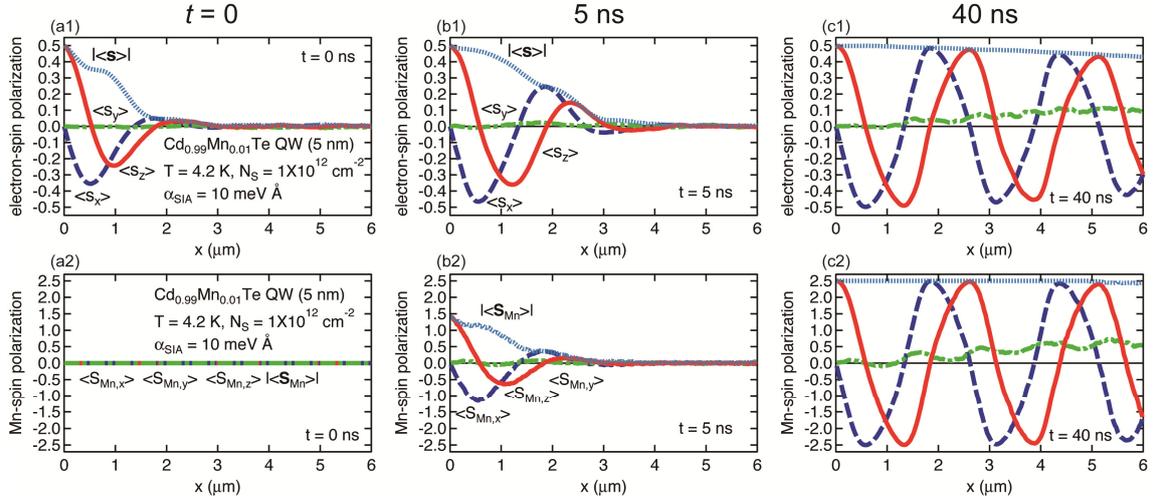

Fig. 3. (Color online) Electron- and Mn-spin polarizations in a 5nm $Cd_{0.99}Mn_{0.01}Te$ quantum well with electron sheet density $N_S = 10^{12}$ cm$^{-2}$, Rashba coefficient $\alpha_{Rashba} = 10$ meVÅ, and in-plane electric field $E_x = -1$ kV/cm as a function of distance $x$ from the source electrode at temperature $T = 4.2$ K. For electron spins, the effective magnetic field due to the dynamically induced Mn-spin polarization and the Rashba field are taken into account. The small $y$-component is caused by the small tilt of the total effective magnetic field.

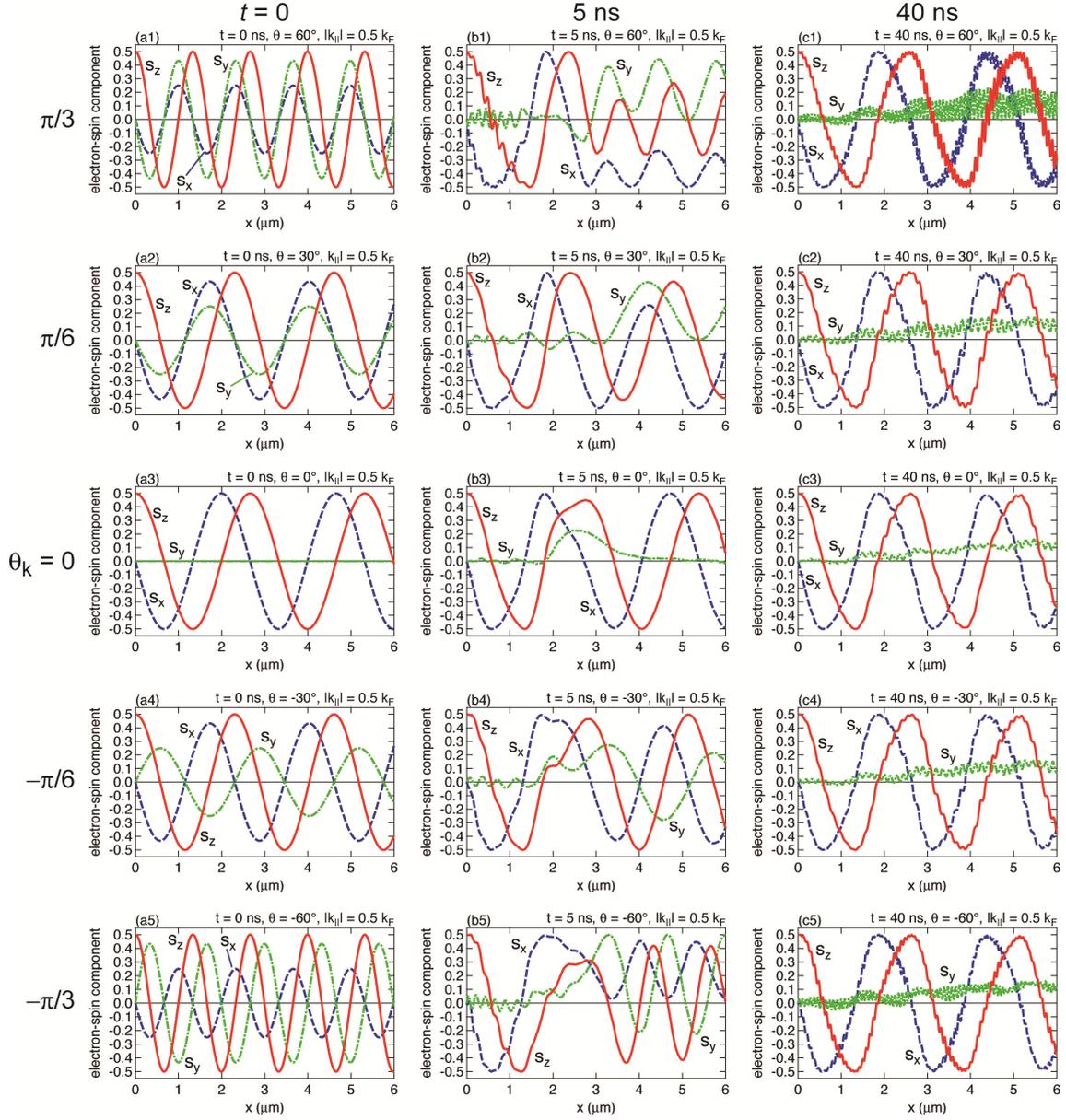

Fig. 4. (Color online) Spin components of an electron with constant 2D wave vector $\mathbf{k}_\| = (k_F/2)\cdot(\cos\theta_k, \sin\theta_k)$ in 5 nm Cd$_{0.99}$Mn$_{0.01}$Te quantum well with electron sheet density $N_S = 10^{12}$ cm$^{-2}$ under the Mn-spin polarization shown in Figs. 3(a2)-3(c2) and the Rashba field of $\alpha_{\text{Rashba}} = 10$ meVÅ at $T = 4.2$ K. Without Mn-spin polarization or for $t = 0$, $\mathbf{s}(x)$ depends on $\theta_k$. Under the polarized Mn spins, at $t = 40$ ns and $x < 1.5$ μm

at $t = 5$ ns, electron spins are synchronized with Mn polarization regardless of $\theta_k$. The fine oscillation prominent for $t = 40$ ns originates from the small difference in direction between the total effective magnetic field and the electron spin.

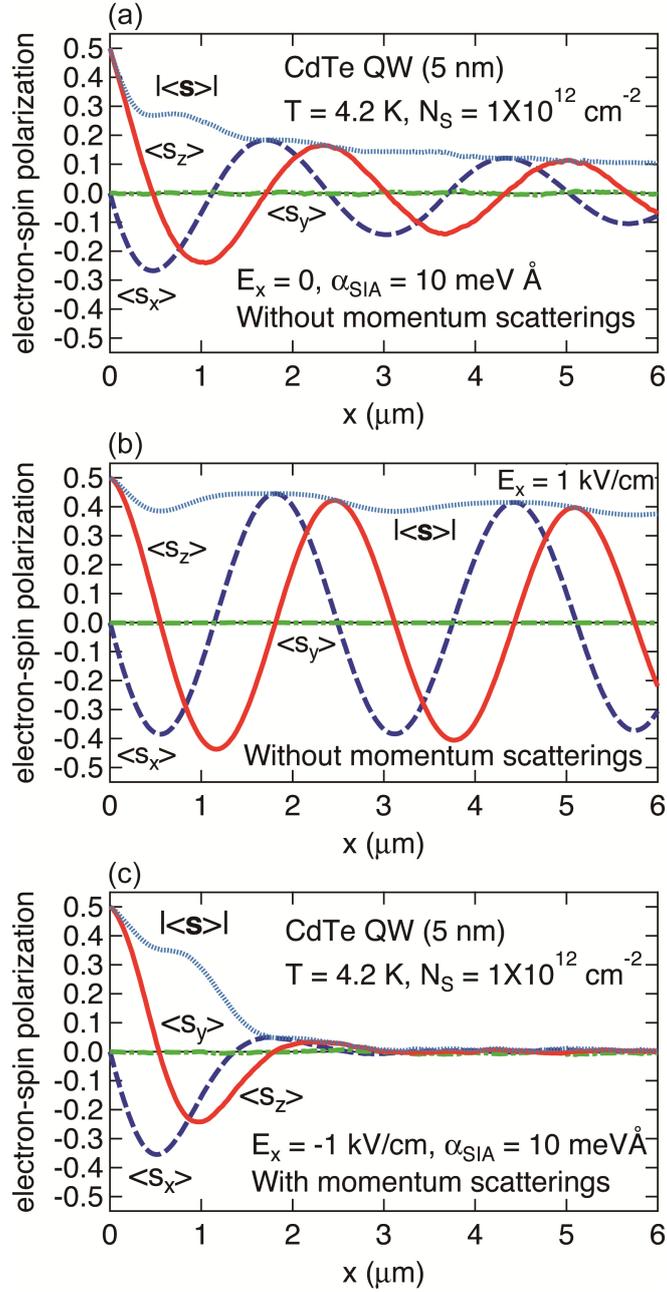

Fig. 5. (Color online) Electron-spin polarization in a 5nm CdTe quantum well as a function of distance $x$ from the ferromagnetic source electrode at (a) $E_x = 0$ without momentum scatterings, (b) $E_x = -1$ kV/cm without momentum scatterings, and (c) $E_x = -1$ kV/cm with momentum scatterings. The electron sheet density is $N_S = 10^{12}$ cm$^{-2}$, the Rashba coefficient $\alpha_{\rm Rashba} = 10$ meVÅ, and temperature $T = 4.2$ K.

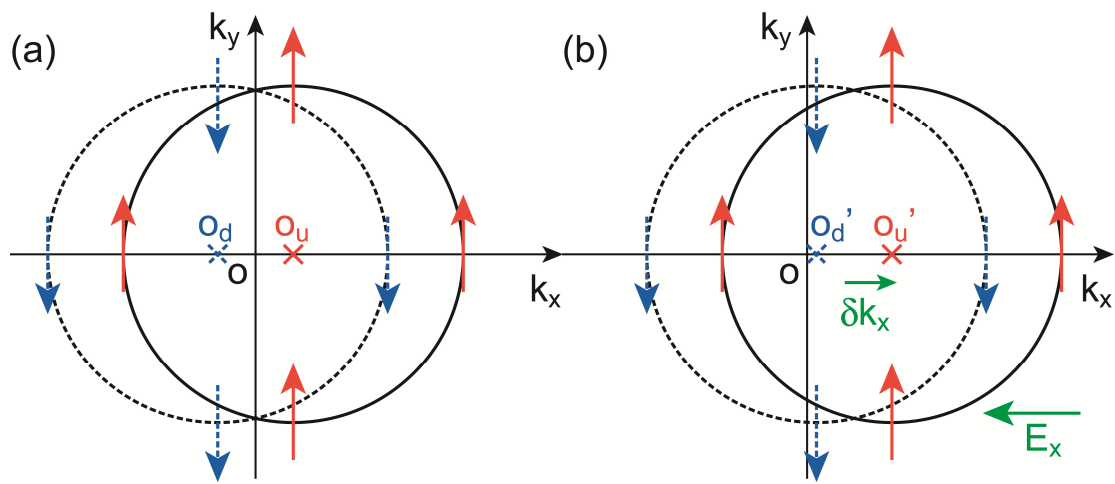

Fig. 6. (Color online) Schematic illustration of electron distribution for spin-up and down states along the $y$-axis under the Rashba effective magnetic field: (a) circular Fermi surfaces for $E_x = 0$ and (b) shifted Fermi circles for $E_x < 0$ in the Ohmic conduction regime. The centers of the Fermi circles for spin-up and spin-down electrons are denoted by $O_u$ and $O_d$ for $E_x = 0$ and by $O_u'$ and $O_d'$ for $E_x \neq 0$, respectively.

Table I. Parameters used in the numerical calculations. Some parameters used to estimate electron scattering rates are not referred in the text.

| Material | $Cd_{0.99}Mn_{0.01}Te$ |
|---|---|
| Lattice constant $a$ | 0.6482 nm (ref. 14) |
| Well thickness $d$ | 5 nm |
| Temperature $T$ | 4.2 K |
| Electron sheet density $N_S$ | $10^{12}$ cm$^{-2}$ |
| In-plane electric field $E_x$ | -1 kV/cm |
| Effective mass $m^*$ | 0.09 $m_0$ (ref. 14) |
| Rashba constant $\alpha_{Rashba}$ | 10 meVÅ |
| s-d coupling constant $N_0 a_{s-d}$ | 220 meV (ref. 7) |
| Density of cation sites $N_0 = 4/a^3$ | 1.468 x $10^{22}$ cm$^{-3}$ |
| Spin lattice relaxation time for Mn $\tau_{Mn}^{lattice}$ | 0.1 ms (4.2 K) |
| Static dielectric constant $\varepsilon_0$ | 10.2 (ref. 14) |
| High frequency dielectric constant $\varepsilon_\infty$ | 7.1 (ref. 14) |
| LO phonon energy $\hbar\omega_{LO}$ | 21.01 meV (ref. 14) |
| Acoustic deformation potential $d_a$ | 9.5 eV (ref. 32) |
| Longitudinal elastic constant $C_L$ | 6.97x$10^{10}$ N/m$^2$ (ref. 32) |
| Impurity scattering time for electrons $\tau_e^{imp}$ | 15 ps |

$m_0$ : electron rest mass